\documentclass{aa}  

\usepackage[colorlinks=true]{hyperref}
\hypersetup{linkcolor=blue, citecolor=blue}
\usepackage{graphicx}
\usepackage{txfonts}
\begin{document} 

\newcommand{\rskal}[1] { {\color{magenta} #1} }
\newcommand{\Btot}{$\vec{B}_{tot}$}
\newcommand{\VA}{$V_{A}$}
\newcommand{\MA}{$\mathcal{M_{A}}$}
\newcommand{\MS}{$\mathcal{M}_{s}$}
\newcommand{\dL}{$\delta \mathcal{L}$}
\newcommand{\dLperp}{$\delta \mathcal{L_{\perp}}$}
\newcommand{\dLparal}{$\delta \mathcal{L_{\parallel}}$}
\newcommand{\dvparal}{$\delta v_{\parallel}$}
\newcommand{\dvparall}{$\delta v_{\parallel, \ell}$}
\newcommand{\dvperp}{$\delta v_{\perp}$}
\newcommand{\dvperpl}{$\delta v_{\perp, \ell}$}
\newcommand{\dBparal}{$\delta B_{\parallel}$}
\newcommand{\dBparall}{$\delta B_{\parallel, \ell}$}
\newcommand{\dBperp}{$\delta B_{\perp}$}
\newcommand{\dBperpl}{$\delta B_{\perp, \ell}$}
\newcommand{\Bordered}{$\vec{B}_{0}$}
\newcommand{\Bturb}{$\delta \vec{B}$}
\newcommand{\Lparallel}{$\ell_{\parallel}$}
\newcommand{\Lperp}{$\ell_{\perp}$}
\newcommand{\uparallel}{$u_{\parallel}$}
\newcommand{\uparallell}{$u_{\parallel, \ell}$}
\newcommand{\uperp}{$u_{\perp}$}
\newcommand{\uperpl}{$u_{\perp, \ell}$}
\newcommand{\Tparallel}{$\rm T_{\parallel}$}
\newcommand{\Tperp}{$\rm T_{\perp}$}
\newcommand{\Ekinetic}{$\rm E_{kinetic}$}
\newcommand{\Ecoupling}{$\rm E_{coupling}$}
\newcommand{\Eharmonic}{$\rm E_{harmonic}$}
\newcommand{\CouplingPotential}{$\vec{B}_{0} \cdot \delta \vec{B}$}
\newcommand{\HarmonicPotential}{$\delta B^{2}$}

   \title{Analytic characterization of sub-Alfv\'enic turbulence energetics}

   \subtitle{}
    \titlerunning{Sub-Alfv\'enic turbulence energetics}
        
   \author{R. Skalidis\inst{1}\thanks{skalidis@caltech.edu}, K. Tassis\inst{2} \fnmsep \inst{3}, \and V. Pavlidou\inst{2} \fnmsep \inst{3}
          }
        
     \institute{Owens Valley Radio Observatory, California Institute of Technology, MC 249-17, Pasadena, CA 91125, USA \and
        Department of Physics \& ITCP, University of Crete, GR-70013, Heraklion, Greece
        \and
        Institute of Astrophysics, Foundation for Research and Technology-Hellas, Vasilika Vouton, GR-70013 Heraklion, Greece}
        
   \date{}

 
  \abstract{Magnetohydrodynamic (MHD) turbulence is a cross-field process relevant to many systems. A prerequisite for understanding these systems is to constrain the role of MHD turbulence, and in particular, the energy exchange between kinetic and magnetic forms. The energetics of strongly magnetized and compressible turbulence has so far resisted attempts to understand them. Numerical simulations reveal that kinetic energy can be orders of magnitude higher than fluctuating magnetic energy. We solved this lack-of-balance puzzle by calculating the energetics of compressible and sub-Alfv\'enic turbulence based on the dynamics of coherent cylindrical fluid parcels. Using the MHD Lagrangian, we  proved analytically that the bulk of the magnetic energy transferred to kinetic energy is the energy that is stored in the coupling between the ordered and fluctuating magnetic field. The analytical relations are in strikingly good agreement with numerical data, up to second-order terms.
}


   \maketitle
%

\section{Introduction}

Magnetohydrodynamic (MHD) turbulence is involved in a plethora of physical phenomena \citep{Biskamp_2003, beresnyak_2019, Matthaeus_2011, Matthaeus_2021, mhd_biased_review}. 
The interplay between kinetic and magnetic energy is important for understanding these processes \citep{goldstein_review, ciolek_2006, kirk_2009, oughton_2013, matthaeus_1983, zweibel_mckee_1995, schekochihin_2007, cho_lazarian_2002, federrath_2011}. It is challenging to understand the energy exchange between kinetic and magnetic forms because the MHD equations are nonlinear. For this reason, several assumptions and approximations are usually employed.

A widely employed approximation is the incompressibility of the gas \citep{sridhar_1994, goldreich_1995}, although this is only applicable to a limited number of systems. Compressible MHD turbulence is more complex, and additional energy terms contribute to the energy cascade. One main difference in the energy cascade rate of incompressible and compressible turbulence is that in the latter, the background magnetic field (\Bordered) appears with leading-order terms \citep{Banerjee_2013, andres_2017}. In contrast, the incompressible turbulence energy cascade is dominated by the increments of the magnetic and velocity fluctuations, and \Bordered\ only appears in higher-order statistics \citep{wan_oughton_servidio_matthaeus_2012}. This result motivated the hypothesis that \Bordered\ might also appear in the total (kinetic and magnetic) fluctuating energy of compressible MHD turbulence \citep{andres_2017}, whereas in incompressible turbulence, the total fluctuating energy is dominated by the fluctuating (second-order) kinetic and magnetic energy.

In incompressible and sub-Alfv\'enic turbulence, the fluctuating magnetic energy is completely transferred to kinetic energy, and the volume-averaged quantities are in equilibrium, $\rho \langle u^{2}\rangle/2 \sim \langle \delta B^{2} \rangle / 8\pi$, when turbulence is maintained in a steady state. In contrast, direct numerical simulations of sub-Alfv\'enic and compressible turbulence show that the volume-averaged kinetic energy is much higher than the second-order fluctuating magnetic energy, $\rho \langle u^{2}\rangle/2 \gg \langle \delta B^{2} \rangle / 8\pi$ \citep{heitsch_2001, li_ps_2012a, li_ps_2012b}, and their relative ratio depends on the amplitude of \Bordered\ \citep{andres_2018, Lim_Cho_2020, skalidis_2021_bpos, beattie_2022_energy_balance}. The excess of the kinetic energy suggests that \Bordered\ might provide additional energy to the fluid. 

The role of \Bordered\ in the energetics can be intuitively understood when we decompose the total magnetic field into a background and a fluctuating component. In incompressible turbulence, the fluctuating magnetic energy comes only from the perturbations of the magnetic field, which are of second order. However, in compressible turbulence, the background field appears in the the total fluctuating magnetic energy due to the coupling between the background field and magnetic perturbations (\Bturb). The magnetic coupling, expressed as \CouplingPotential, can only be realized in compressible turbulence \citep{montgomery_1987, Bhattacharjee_1988, Bhattacharjee_1998, fujimura_2009} and is the dominant (first-order) term of the fluctuating magnetic energy.

In sub-Alfv\'enic and compressible turbulence, numerical data show that \CouplingPotential\ stores most of the magnetic energy, and that the kinetic energy approximately reaches equipartition with the fluctuations of the coupling term \citep{skalidis_tassis2021, skalidis_2021_bpos, beattie_2022_energy_balance, beattie_2022_review}. Thus, the magnetic coupling holds the key for understanding the energetics of strongly magnetized and compressible turbulence. However, there is still a lack of first-principle understanding of the role of \CouplingPotential\ in MHD turbulence dynamics and how it contributes to the averaged energetics.  

We present an analytical theory of the role of the coupling potential in the energy exchange of sub-Alfv\'enic and compressible turbulence, which is encountered in systems such as tokamaks \citep{strauss_1976, tokamaks_1977, zocco_2011}, in the interstellar medium \citep{mouschovias_2006, panopoulou_2015, panopoulou_2016, planck_collaboration_2016, skalidis_2022}, and the Sun \citep{verdini_velli_2007, tenerani_velli_2017, kasper_2021, zank_2022}. We write the Lagrangian of coherent flux structures \citep{crowley_2022}, which allows us to approximate turbulence properties in a deterministic manner, and calculate analytically the energy exchange between kinetic and magnetic forms as a function of the Alfv\'enic Mach number (\MA). We find remarkable agreement between the analytically calculated energetics and numerical data. We conclude that the majority of the fluctuating magnetic energy transferred to kinetic energy is provided by the coupling between the background and the fluctuating magnetic field. 

\begin{figure}[ht]
    \centering \includegraphics[width=1.0\hsize]{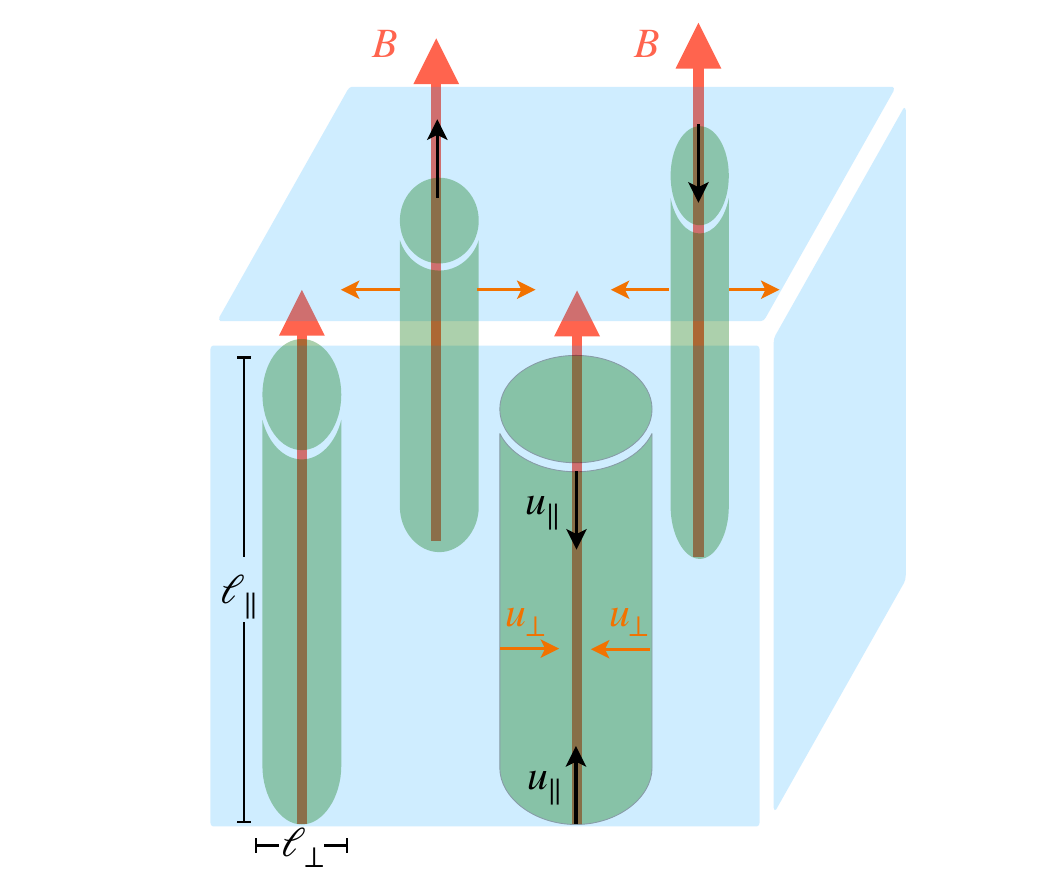}
    \caption{Magnetized fluid consisting of multiple coherent cylindrical fluid parcels. Red arrows show the initial magnetic field morphology. Untwisted fluid parcels are elongated, \Lparallel\ $\gg$ \Lperp, and their motion is longitudinal along or perpendicular to \Bordered. In sub-Alfv\'enic turbulence, the motion of these fluid parcels can be decomposed into two independent velocity components, parallel (black arrows) and perpendicular (orange arrows) to \Bordered.}
    \label{fig:flux_tubes}
\end{figure}

\section{Model}

We considered a turbulent fluid  characterized by the commonly employed properties: 1) spatial homogeneity, 2) infinite magnetic and kinetic Reynolds number, and 3) time stationarity. We considered that the fluid consists of coherent flux tubes \citep[e.g., Fig. 1 in][]{Banerjee_2013} (or fluid parcels) with coordinates $\left( r(t), \phi(t), z(t) \right)$, as shown in Fig.~\ref{fig:flux_tubes}. Cylindrical coordinates are motivated by studies showing that the properties of strongly magnetized turbulence are axially symmetric, with \Bordered\ being the axis of symmetry \citep{goldreich_1995, maron_2001}. We assumed the following initial conditions: 1) uniform temperature, 2) uniform density, 3) no bulk velocity, 4) uniform static magnetic field (\Bordered\ = $B_{0} \hat{z}$), and 5) no self-gravity. We henceforth adopt the following notation: $z = \ell_{\parallel}$ and $r = \ell_{\perp}$, where $\ell_{\parallel}$ and $\ell_{\perp}$ denote parcel sizes parallel and perpendicular to \Bordered\ , respectively. 

We perturbed the magnetic field of a coherent fluid structure with a length scale $\ell=\sqrt{\ell^{2}_{\parallel} + \ell_{\perp}^{2}}$ by $\delta \vec{B}_{\ell}$ such that $|\vec{B}_{0}| \gg |\delta \vec{B}_{\ell}|$, which applies to sub-Alfvénic turbulence. Magnetic perturbations tend to redistribute the magnetic flux within a fluid. For ideal-MHD (flux-freezing) conditions, the magnetic flux is preserved. Thus, the surface of the perturbed fluid parcel ($\vec{S_{\ell}}$) follows the magnetic field lines. The motion of the field lines, and hence of $\vec{S_{\ell}}$, can be either parallel or perpendicular to \Bordered\ (Fig.~\ref{fig:flux_tubes}): 1) Squeezing and stretching of $\vec{S_{\ell}}$ along \Bordered\ leads to parallel motions, $\dot{\ell}_{\parallel} \neq 0$. 2) Fluctuations of $\ell_{\perp}$ lead to perpendicular motions, $\dot{\ell}_{\perp} \neq 0 $. Finally, 3) twisting leads to rotational motions, $\dot{\phi} \neq 0 $. This naturally defines $\ell_{\parallel}$ and $\ell_{\perp}$ as the coherence lengths of the perturbed volume parallel and perpendicular to \Bordered\ , respectively. We focused on large scales since coherent structures are prominent there \citep{de_giorgio_2017}. We invoke as a boundary condition a local environment beyond $\ell$ (pressure wall).

The flux freezing theorem is
\begin{equation}
    \label{eq:alfven_theorem}
    \frac{d \vec{B_{\ell}}}{d t} \cdot \vec{S_{\ell}} = - \vec{B_{\ell}} \cdot \frac{d \vec{S_{\ell}}}{d t}.
\end{equation}
The cross section of the coherent volume perpendicular and parallel to \Bordered\ is $\vec{S}_{\perp, \ell} = 2\pi \ell_{\perp} \ell_{\parallel} \hat{r}$, and $\vec{S}_{\parallel, \ell} = \pi \ell^{2}_{\perp} \hat{z}$ , respectively. The cross section related to the rotational motion is $\vec{S}_{\phi, \ell} = \ell_{\parallel} \ell_{\perp} \hat{\phi}$. The total magnetic field in cylindrical coordinates can be expressed as $\vec{B_{\ell}} = \delta B_{\perp r, \ell} \hat{r} + \delta B_{\perp \phi, \ell} \hat{\phi} + \left(B_{0}+\delta B_{\parallel, \ell}\right) \hat{z}$. From Eq.~\ref{eq:alfven_theorem}, we obtain that when $|\vec{B}_{0}| \gg |\delta \vec{B}|$, magnetic perturbations along $\vec{S}_{\parallel, \ell}$ are associated with a longitudinal motion such that
\begin{equation}
    \label{eq:u_perp_generalized}
     u_{\perp r, \ell} \equiv \dot{\ell}_{\perp} (t) = 
     - \frac{\delta \dot{B}_{\parallel, \ell}(t)}{2 B_{0}+\delta B_{\parallel, \ell}} \ell_{\perp}(t) \approx - \frac{\delta \dot{B}_{\parallel, \ell}(t)}{2 B_{0}} \ell_{\perp, 0},
\end{equation}
where we have considered that the initial dimension of the perturbed volume $\ell_{\perp, 0}$ is much larger than its perturbations. Along $\vec{S}_{\perp, \ell}$ , we find that
 \begin{equation}
 \label{eq:u_parallel_generalized}
    u_{\parallel, \ell} \equiv \dot{\ell}_{\parallel}(t) = -\left ( \frac{\delta \dot{B}_{\perp r, \ell}(t)}{\delta B_{\perp r, \ell}(t)} - \frac{\delta \dot{B}_{\parallel, \ell}(t)}{2 B_{0}} \right) \ell_{\parallel}(t) \approx - \frac{\delta \dot{B}_{\perp r, \ell}(t)}{\delta B_{\perp r, \ell}(t)} \ell_{\parallel} (t),
\end{equation}
while the azimuthal velocity along $\vec{S}_{\phi, \ell}$  is 
\begin{equation}
  \label{eq:omega_generalized_total}
  u_{\perp \phi, \ell} \equiv \dot{\phi}(t) \ell_{\perp}(t) \approx - \left(\frac{\delta \dot{B}_{\perp r, \ell}(t)}{\delta B_{\perp r, \ell}(t)} - \frac{\delta \dot{B}_{\perp \phi, \ell}(t)}{\delta B_{\perp \phi, \ell}(t)} \right) \ell_{\perp}(t).
\end{equation}
As a result of assuming $|\vec{B}_{0}| \gg |\delta \vec{B}|$, we have obtained that parallel and perpendicular motions are decoupled. The coupling of parallel and perpendicular motions becomes inevitable when $|\vec{B}_{0}| \sim |\delta \vec{B}|$ (Eq.~\ref{eq:u_parallel_generalized}).

In sub-Alfv\'enic turbulence, magnetic tension dominates magnetic pressure \citep{passot_2003}. The high tension suppresses transverse oscillations due to the strong restoring torques. Thus, twisting would have minimum contribution to the dynamics
\citep[e.g.,][]{longcope_1997} and motions would be mostly longitudinal ($\dot{\phi} \, , \delta B_{\perp \phi, \ell} \approx 0$). Since $u_{\perp \phi, \ell} \rightarrow 0 $, then due to Eqs.~\ref{eq:u_parallel_generalized}, and \ref{eq:omega_generalized_total}, $\ell_{\parallel} \gg \ell_{\perp}$, which implies that untwisted coherent structures are stretched toward the \Bordered\ axis, which is consistent with the anisotropic properties of sub-Alfvénic turbulence \citep{shebalin_1983, higdon_1984, oughton_1994,
sridhar_1994, goldreich_1995, oughton_2013, oughton_2020, cho_lazarian_2003, Yang_2018, makwana_2020, gan_2022}. 

For untwisted fluid parcels, the perpendicular component of the magnetic fluctuations has a dominant radial component such that $\delta B_{\perp, \ell} \approx \delta B_{\perp r, \ell} $. From Eqs.~\ref{eq:u_perp_generalized} and \ref{eq:u_parallel_generalized}, we derive 
\begin{align}
    \label{eq:db_parall_lperp}
    \delta B_{\parallel, \ell}(t) & \propto -B_{0} \log \ell_{\perp}(t) , \\
    \label{eq:db_perp_lparal}
    \delta B_{\perp, \ell}(t) & \propto \ell_{\parallel}^{-1}(t).
\end{align}
The difference in the scaling is due to the Lorenz force by \Bordered, which affects perpendicular motions, while it has no effect on parallel motions.

\section{MHD Lagrangian of coherent structures}

We write the Lagrangian for the perturbed volume. We place the reference frame at the center of mass of the target volume, hence there is no bulk velocity term in the Lagrangian. Therefore, all the velocity components are due to internal motions induced by magnetic perturbations. We focus on low plasma-beta fluids\footnote{The relative ratio of the thermal and magnetic pressure is the plasma beta, which is defined as $\beta = 2  \mathcal{M}^{2}_\mathcal{A}/\mathcal{M}^{2}_\mathcal{S}$.}, which for sub-Alfv\'enic turbulence corresponds to high sonic Mach numbers (\MS). The perturbed Lagrangian \citep{Newcomb1962, andreussi_lagrangian, kulsrud_book} of the coherent cylindrical fluid parcel, with surface $\vec{S_{\ell}}$, can be split into a parallel and a perpendicular term (Appendix~\ref{sec:lagrangian}),
\begin{equation}
    \label{eq:total_lagrangian}
    \delta \mathcal{L} = \overbrace{\left ( \frac{1}{2}\rho  \dot{\ell}_{\parallel}^{2} - \frac{\delta B_{\perp, \ell}^{2}}{8\pi}  \right ) }^{\delta \mathcal{L}_{\perp}} + 
    \overbrace{\left ( \frac{1}{2}\rho \dot{\ell}_{\perp}^{2} - \frac{B_{0} \delta B_{\parallel, \ell}}{4 \pi} - \frac{\delta B_{\parallel, \ell}^{2}}{8\pi} \right )}^{\delta \mathcal{L}_{\parallel}}.
\end{equation}
Due to Eqs.~\ref{eq:u_perp_generalized} and \ref{eq:u_parallel_generalized}, \dBparall, and \dBperpl\ are generalized coordinates of $\delta \mathcal{L}$ and $\ell_{\parallel}(t) = C/\delta B_{\perp, \ell}(t)$ (Eq.~\ref{eq:db_perp_lparal}), where $C$ is a constant determined from the initial conditions. With this expression, we eliminate \Lparallel\ from the Lagrangian, which up to second-order terms is separable into a parallel and a perpendicular part, and is analytically solvable,
\begin{align}
    \label{eq:perturbed_lagrangian_simplified_perp}
    & \delta \mathcal{L}_{\perp}\left(\delta B_{\perp, \ell}, \delta \dot{B}_{\perp, \ell}\right) \approx \frac{1}{2}\rho C^{2} \frac{\delta \dot{B}_{\perp, \ell}^{2}}{\delta B_{\perp, \ell}^{4}}  - \frac{\delta B_{\perp, \ell}^{2}}{8\pi}, \\
    &
    \label{eq:perturbed_lagrangian_simplified_parallel}
    \delta \mathcal{L}_{\parallel}\left(\delta B_{\parallel, \ell}, \delta \dot{B}_{\parallel, \ell}\right) \approx \frac{1}{8}\rho \frac{\delta \dot{B}_{\parallel, \ell}^{2}}{B_{0}^{2}} \ell_{\perp, 0}^{2} - \frac{B_{0} \delta B_{\parallel, \ell}}{4 \pi} - \frac{\delta B_{\parallel, \ell}^{2}}{8\pi}.
\end{align}
We solve the Euler-Lagrange equations for \dLparal\ (Appendix~\ref{sec:solutions_dlparal}) and \dLperp\ (Appendix~\ref{sec:solutions_dlperp}) and derive the analytical solutions of the velocity (\uparallell (t), \uperpl (t) ) and magnetic fluctuations (\dBparal$_{, \ell}(t)$, \dBperp$_{, \ell}(t)$) of $\vec{S_{\ell}}$. We find that \dBparall\ $\sim t^{2}$, \uperpl\ $\sim t$, and \dBperpl\ $\sim t^{-1}$, while \uparallell\ is set by the initial conditions (\textbf{free streaming of the gas}). We used these analytical solutions in order to calculate the averaged energetics of a strongly magnetized and compressible fluid. 

\section{Energetics}
\label{sec:energetics}

The total energy of fully developed turbulence is stationary because energy diffusion is balanced by injection. Time stationarity enables us to approximate turbulence energetics with the leading-order solutions that we obtained because our approximations preserve time symmetry, and thus energy is conserved. The statistical properties of large-scale coherent structures accurately approximate the volume-averaged turbulent statistical properties. Thus, for an ergodic fluid \citep{monin_ergodicity, galanti_2004_ergodicity}, averaging the turbulent statistical properties over the volume of the fluid at a given time step $\left( \langle f \rangle_{\mathcal{V}} = \int_{\mathcal{V}} f \right)$ is approximately equivalent to averaging over multiple realizations of a typical large-scale coherent structure $\left( \langle f_{\ell} \rangle_{\mathcal{T}} = \int_{\mathcal{T}} f_{\ell} \right)$, hence  $ \langle f \rangle_{\mathcal{V}} \sim  \langle f_{\ell} \rangle_{\mathcal{T}}$, where $f$ denotes an energy term, and $\mathcal{T}$ corresponds to the coherent structure crossing time.
We next analytically compute the $\langle f_{\ell} \rangle_{\mathcal{T}}$ energy contribution of each Lagrangian term (Eq.~\ref{eq:total_lagrangian}) and their relative ratios. Since  coherent cylindrical parcels are characterized by two different coherence lengths \Lparallel and \Lperp, they also have two different crossing times: $\mathcal{T}_{\parallel}$ and $\mathcal{T}_{\perp}$ , respectively. We compare the $\langle f_{\ell} \rangle_{\mathcal{T}}$ analytical energy ratios with the $\langle f \rangle_{\mathcal{V}}$ numerical values. The numerical results correspond to simulations of ideal isothermal MHD turbulence without self-gravity, and turbulence is maintained in a quasi-static state by injecting energy with an external forcing mechanism \citep{beattie_2022_energy_balance}. These simulation are forced with a mixture of compressible and incompressible modes, but the driving modes do not affect the energetics of sub-Alfv\'enic and compressible turbulence \citep{skalidis_2021_bpos}.

\subsection{Kinetic energy}

The total averaged kinetic energy (\Ekinetic) of the coherent fluid parcel with scale $\ell$ is 
\begin{equation}
    \label{eq:average_kinetic_energy}
    \rm{E_{kinetic}} \equiv 
    \frac{1}{2} \rho \langle u_{\ell}^{2} \rangle_{\mathcal{T}} = 
    \frac{1}{2} \rho \left ( \langle u_{\perp, \ell}^{2} \rangle_{\mathcal{T}_{\perp}} + \langle u_{\parallel, \ell}^{2} \rangle_{\mathcal{T}_{\parallel}} \right) \approx \frac{B_{0} \delta B_{\parallel, \max}}{6 \pi} + \frac{\delta B_{\perp, \max}^{2}}{8 \pi}
.\end{equation}
The kinetic energy is dominated to first order by \uperpl. Thus, the average Alfvénic Mach number to first order is 
\begin{equation}
    \label{eq:alfven_mach_number}
   \mathcal{M_{A}} \equiv
   \frac{\sqrt{\langle u_{\ell}^{2} \rangle_{\mathcal{T}}}}{V_{A}}
   \approx \sqrt{\frac{4 \delta B_{\parallel, \max}}{3 B_{0}}}.
\end{equation}

\subsection{Harmonic potential}

From Eqs.~\ref{eq:delta_B_parall_solution} and~\ref{eq:deltaB_perp_solution}, we find that $\langle \delta B_{\parallel, \ell}^{2} \rangle_{\mathcal{T_{\perp}}}  = 7\delta B_{\parallel, \max}^{2}/15$, and  $\langle \delta B_{\perp, \ell}^{2} \rangle_\mathcal{T_{\parallel}}  =  \delta B_{\perp, \max}^{2}/2$. The total time-averaged harmonic potential energy (\Eharmonic) density is equal to
\begin{equation}
    \rm{E_{harmonic}} \equiv \frac{\langle \delta B_{\ell}^{2} \rangle_\mathcal{T}}{8\pi} \approx \frac{\delta B_{\parallel, \rm max}^{2}}{8\pi} \left ( \frac{7}{15} + \frac{\zeta^{2}(\mathcal{M}_{A})}{2} \right ),
\end{equation}
where $\zeta = \delta B_{\perp, \max}/\delta B_{\parallel, \max}$. Sub-Alfvénic turbulence is anisotropic \citep{shebalin_1983, higdon_1984, oughton_1994, goldreich_1995}, with the anisotropy between \dBperp\ and \dBparal\ depending on \MA\ \citep{beattie_2020}. To account for this property, we assumed that $\zeta$ 
is a function of \MA. When $\mathcal{M_{A}} \rightarrow 0$, \Bordered\ suppresses any bending of the magnetic field lines with the amplitude of \dBparal\ being larger than that of \dBperp\ \citep{beattie_2020}, hence $\zeta \rightarrow 0$; this is also a consequence of $\vec{\nabla} \cdot \vec{B} = 0$ for anisotropic fluid parcels with  \Lparallel\ $\gg$ \Lperp. For $\mathcal{M_{A}} \rightarrow 1$, fluctuations tend to become more isotropic, and hence $\zeta \rightarrow \sqrt{2}$. These limiting behaviors are consistent with numerical simulations \citep{beattie_2020, beattie_2022_energy_balance}.

\subsection{Coupling potential}

According to Eq.~\ref{eq:average_kinetic_energy}, \CouplingPotential\
contributes to \Ekinetic\ since
\begin{equation}
\label{eq:average_coupling_term}
    \rm{E_{coupling}} \equiv \frac{B_0 \sqrt{\langle \delta B_{\parallel, \ell}^{2} \rangle_{\mathcal{T_{\perp}}}}}{4\pi} 
    \approx \frac{B_0 \delta B_{\parallel, \max}}{6\pi} 
    \approx \rm{E_{kinetic}},
\end{equation}
to first order.
This equation demonstrates that the energy stored in the coupling potential is in equipartition with the averaged kinetic energy when turbulence is sub-Alfvénic. 

\subsection{Energetics ratios}

The \Ekinetic$/$\Ecoupling\ ratio is
\begin{equation}
    \label{eq:ratio_kinetic_coupling}
    \frac{\rm E_{kinetic}}{\rm E_{coupling}} \approx 1 + \frac{9}{16} \mathcal{M}^{2}_{A} \zeta^{2}(\mathcal{M_{A}}).
\end{equation}
For $\mathcal{M_{A}} \rightarrow 0$, \Ecoupling\ $\approx$ \Ekinetic, while for $\mathcal{M_{A}} \rightarrow 1$, \Ekinetic\ $\gtrsim$ \Ecoupling. \Ekinetic\ becomes higher than \Ecoupling\ because  \uparallel$_{, \ell}$ contributes more to \Ekinetic\ as \MA\ increases. When $\mathcal{M_{A}} \rightarrow 1$,  $\zeta\approx\sqrt{2}$ , so that the \Ekinetic/\Ecoupling\ ratio in trans-Alfvénic turbulence scales as
\begin{equation}
    \label{eq:average_kinetic_coupling}
    \frac{\rm E_{kinetic}}{\rm E_{coupling}} \approx 1 + \frac{9}{8} \mathcal{M_{A}}^{2}.
\end{equation}

Regarding the \Eharmonic/\Ecoupling\ ratio, we find that
\begin{equation}
    \label{eq:ratio_coupling_harmonic}
     \frac{\rm E_{harmonic}}{\rm E_{coupling}} \approx \frac{3}{8} \sqrt{\frac{15}{7}} \mathcal{M}^{2}_{\mathcal{A}} \left(\frac{7}{15}+\frac{\zeta^{2}(\mathcal{M_{A}})}{2} \right),
\end{equation}
which for the two limiting cases of $\zeta$ (\MA) becomes 
\begin{equation}
    \label{eq:average_harmonic_potential}
    \frac{\rm E_{harmonic}}{\rm E_{coupling}} \approx
    \left\{\begin{array}{l}
    0.25~\mathcal{M}^{2}_{\mathcal{A}},~\mathcal{M_{A}} \rightarrow 0 \\ \\
    0.80~\mathcal{M}^{2}_{\mathcal{A}},~\mathcal{M_{A}} \rightarrow 1
    \end{array}\right. .
\end{equation}

\subsection{Comparison between analytical and numerical results} 
\label{sec:analytical_numerical}

\begin{figure}
    \centering
    \includegraphics[width=\hsize]{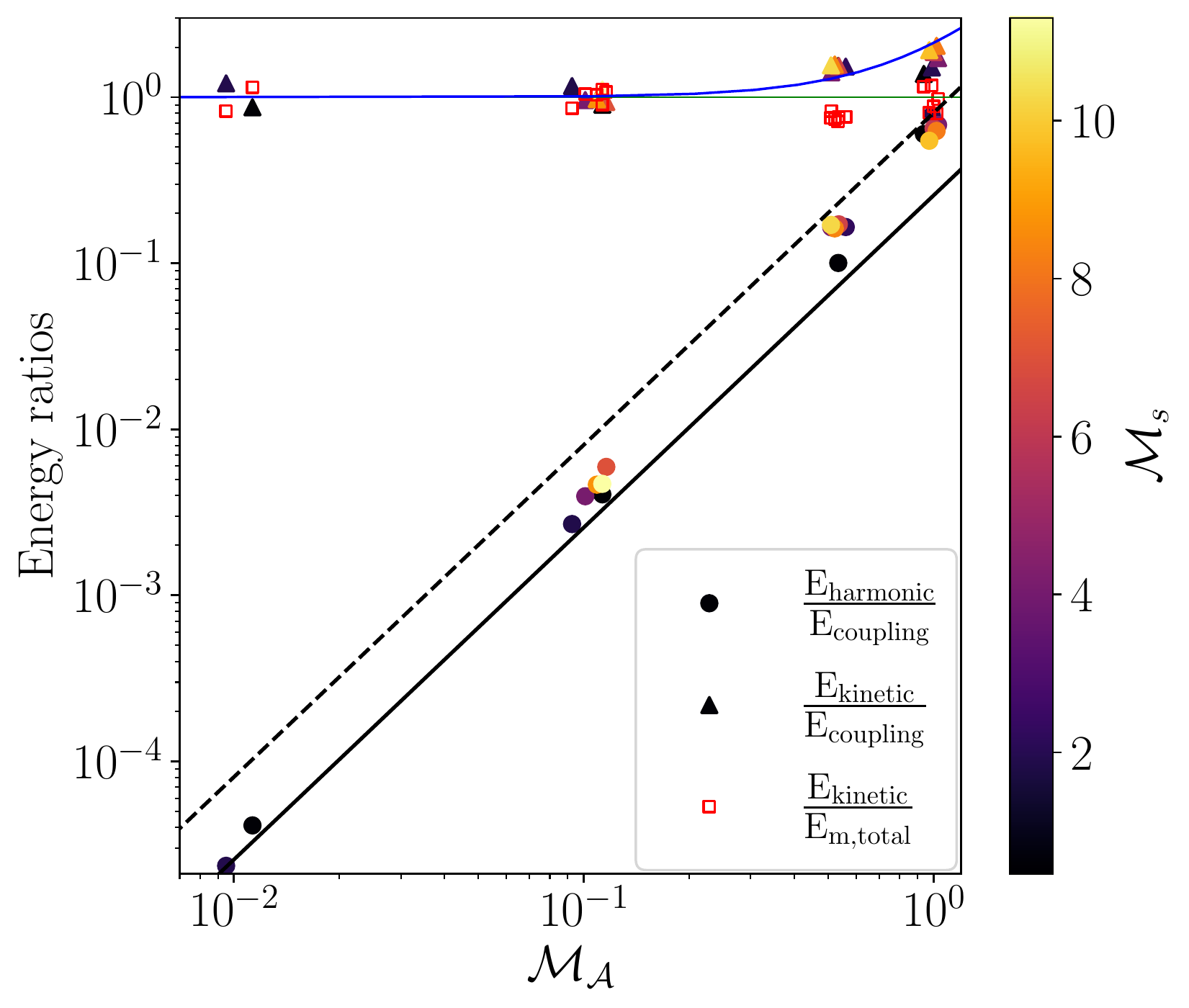}
    \caption{Comparison between analytical and numerical results. The solid and dashed thick black lines correspond to the \Eharmonic/\Ecoupling ratio obtained analytically for $\mathcal{M_{A}} \rightarrow 0 $ ($\zeta=0$) and $\mathcal{M_{A}}  \rightarrow 1 $ ($\zeta=\sqrt{2}$), respectively. Numerical data are shown with colored dots. The blue line corresponds to the analytically obtained \Ekinetic/\Ecoupling\ ratio, while colored triangles show the same quantities calculated from numerical data. The red boxes correspond to \Ekinetic$/\rm E_{\rm m, total}$. The thin green line shows the energy terms in equipartition. The color bar shows the sonic Mach number ($\mathcal{M}_{s}$) of the simulations.}
    \label{fig:energetics_ratio}
\end{figure}

In Fig.~\ref{fig:energetics_ratio} we compare the analytically calculated energy ratios with numerical results from the literature  \citep{beattie_2022_energy_balance}. The lines correspond to the analytical relations for \Eharmonic/\Ecoupling\ (Eq.~\ref{eq:average_harmonic_potential}) and \Ekinetic$/$\Eharmonic\ (Eq.~\ref{eq:average_kinetic_coupling}), while the colored points correspond to the numerical values. The numerical data behave as predicted by the analytical relations. The scatter of triangles increases at higher \MA\  because thermal pressure starts becoming important there, and hence the contribution of thermal motions to the kinetic energy increases. In the limit of \MA\ $\ll 1$, thermal pressure is subdominant and $\beta \rightarrow 0$. For \MA\ $= 1$, we obtain that $\beta \rightarrow 0$ when \MS\ $\gg 1$, while when \MS\ $\lesssim 1$, $\beta \rightarrow 1$. Thus, for trans-Alvf\'enic turbulence, thermal pressure becomes important only for low \MS, while at high \MS\ , it has a minor contribution to the energetics. In our calculations, we neglected thermal pressure, and for this reason, at \MA\ = 1, triangles are consistent with the analytical ratio of \Ekinetic/\Ecoupling\ (blue line) when \MS\ $\geq 2$, while at lower \MS, the deviation between numerical and analytical results increases because $\beta$, hence the relative contribution of thermal pressure, increases. For \MA\ < 1, $\beta \ll 1,$ and for this reason, the numerical data agree perfectly with the analytical ratio (blue line). Finally, when we account for the contribution from both \CouplingPotential\ and \HarmonicPotential, the total energy stored in magnetic fluctuations $\left( \rm E_{m, total} = E_{coupling} + E_{harmonic} \right)$ is very close to equipartition with kinetic energy, as shown by the red boxes.

\section{Discussion and conclusions}

Analytical calculations of strongly magnetized and compressible (isothermal) turbulence show that \Bordered\ appears in the energy cascade with leading-order terms \citep{Banerjee_2013, andres_2018}. This is in striking contrast to incompressible turbulence, where \Bordered\ appears in higher-order terms \citep{wan_oughton_servidio_matthaeus_2012}. In the formalism presented here, the incompressible limit is approximated when \CouplingPotential\ = 0. In this case, \Bordered\ does not appear in the dominant Lagrangian terms, hence the averaged kinetic energy would scale linearly with the fluctuating magnetic energy, $\delta u_{\ell} \sim \delta B_{\ell}$ (or equivalently, $\mathcal{M_{A}} \sim \delta B_{\ell}$). However, in agreement with previous works \citep{wan_oughton_servidio_matthaeus_2012}, our formalism shows that \Bordered\ appears in the energetics of incompressible turbulence because of the coupling between \dBparall and \dBperpl\ (Eqs.~\ref{eq:u_parallel_generalized}) when higher-order terms are considered.

For sub-Alfv\'enic and compressible turbulence, we find that  \CouplingPotential\ is the leading term in the dynamics, and as a result, the scaling between velocity and magnetic fluctuations becomes $\delta u_{\ell} \sim \sqrt{B_{0} \delta B_{\ell}}$, or equivalently, $\mathcal{M}_{\mathcal{A}} \sim \delta u_{\ell}/V_{A} \sim \sqrt{\delta B_{\ell}/B_{0}}$, which is supported by numerical data \citep{beattie_2020}. In compressible and strongly magnetized turbulence, compression and dilatation of the gas locally changes the energy cascade rate \citep{Banerjee_2013}. These local energy fluctuations can only be realized in compressible turbulence and might be related to the fluctuations of the \CouplingPotential\ potential. Our analytical results prove that the total averaged magnetic energy transferred to kinetic is equal to $\left ( 2 B_{0} \sqrt{\langle \delta B_{\parallel}^{2} \rangle} + \langle \delta B^{2} \rangle  \right) /8\pi$. 

The consistency between our analytical relations and numerical data is remarkable. It is not the first time that simple analytical arguments agree quantitatively with numerical simulations of nonlinear problems \citep[e.g.,][]{mouschovias_2011}. However, an analytical theory is always advantageous because it allows us to achieve a deeper understanding of complex problems. For this reason, the formalism we presented might offer new insights into the energetics of strongly magnetized and compressible turbulence. We hope that it motivates future works about the role of magnetic couplings in the energy cascade.

\bibliographystyle{aa}
\bibliography{bibliography}

\begin{acknowledgements}
We would like to thank the anonymous referee for reviewing our manuscript. We are grateful to T. Ch. Mouschovias, and P. F. Hopkins for stimulating discussions. We also thank E. N. Economou, V. Pelgrims, E. Ntormousi, A. Tsouros, and I. Komis for useful suggestions on the manuscript. This work was supported by NSF grant AST-2109127. We acknowledge support by the European Research Council under the European Union's Horizon 2020 research and innovation programme, grant agreement No. 771282 (RS and KT); by the Hellenic Foundation for Research and Innovation under the “First Call for H.F.R.I. Research Projects to support Faculty members and Researchers and the procurement of high-cost research equipment grant”,  Project 1552 CIRCE (VP); and from the  Foundation of Research and Technology - Hellas Synergy Grants Program (project MagMASim, VP, and project POLAR, KT). 
\end{acknowledgements}

\begin{appendix}

\section{Lagrangian of coherent cylindrical parcels}
\label{sec:lagrangian}

We used the MHD Lagrangian as derived by Newcomb \citep{Newcomb1962} for isothermal magnetized fluids. The total MHD Lagrangian is the sum of the kinetic and the total potential energy of all the fluid elements within a volume $\mathcal{V}$,
\begin{equation}
    \mathcal{L} = \int_{\mathcal{V}} \left ( \frac{1}{2}\rho u^{2} - P_{s} - \frac{B^{2}}{8 \pi} - \frac{1}{2}\rho \Phi \right ),
\end{equation}
where $P_{s}$ is the thermal pressure, and $\Phi$ is the gravitational potential. The equation of motion for magnetized turbulent fluids is obtained from the stationary-action principle $\left( \delta \int_{t_{1}}^{t_{2}} dt \mathcal{L} = 0 \right)$. We focused on fluids in which magnetic pressure dominates thermal pressure $\left( B^{2}/8\pi \gg P_{s} \right)$, and we ignored self-gravity, $\Phi = 0$. Therefore, the dominant potential term in the Lagrangian is magnetic pressure. When we consider that the ensemble of the fluid elements moves as a coherent cylinder (Fig.~\ref{fig:flux_tubes}), then the integration of the Lagrangian takes place within the volume of the cylinder. In this case, the integrated Lagrangian terms correspond to the kinetic and the magnetic energy of the cylinder. The perturbed Lagrangian of a cylinder is (using Eqs.~\ref{eq:u_perp_generalized}, \ref{eq:u_parallel_generalized}, and \ref{eq:omega_generalized_total})
    
    \begin{equation}
        \begin{split}
         \delta \mathcal{L} = \frac{1}{2} \rho \ell^{2} \left( \frac{\delta \dot{B}_{\parallel}^{2}}{4 B_{0}^{2}} +  \frac{\delta \dot{B}_{\perp r}^{2}}{\delta B_{\perp r}^{2}} \right) + \frac{1}{2} \rho \ell_{\perp}^{2} \left(  \frac{\delta \dot{B}_{\perp \phi}^{2}}{\delta B_{\perp \phi}^{2}} -   
     2 \frac{\delta \dot{B}_{\perp r}}{\delta B_{\perp r}}\frac{\delta \dot{B}_{\perp \phi}}{\delta B_{\perp \phi}}  \right) \\ - \frac{1}{2} \rho \ell_{\parallel}^{2} \frac{\delta \dot{B}_{\perp r}}{\delta B_{\perp r}}\frac{\delta \dot{B}_{\parallel}}{B_{0}} - \left ( \frac{\delta B_{\parallel}^{2}}{8 \pi} + \frac{\delta B_{\perp r}^{2}}{8 \pi} + \frac{\delta B_{\perp \phi}^{2}}{8 \pi} \right) - \frac{B_{0} \delta B_{\parallel}}{4 \pi}. 
        \end{split}
    \end{equation}

For untwisted cylinders, all terms containing a $\phi$ component are zero. Then, the Lagrangian contains only the parallel and the perpendicular (longitudinal) components, which are generally coupled due to the $\delta \dot{B}_{\perp r} \delta \dot{B}_{\parallel}/(\delta B_{\perp r}B_{0}) \ell_{\parallel}^{2}$ term. For sub-Alfv\'enic turbulence, this is a higher-order term because |\Bordered| $\gg$ |\Bturb|. By keeping the dominant (second-order) terms, we derived the total perturbed Lagrangian of a coherent cylindrical structure, which to leading order, can be expressed as the sum of two independent parts (parallel and perpendicular to the background magnetic field, Eqs.~\ref{eq:perturbed_lagrangian_simplified_perp} and \ref{eq:perturbed_lagrangian_simplified_parallel}).

\section{Solutions of \dLparal}
\label{sec:solutions_dlparal}

From the Euler-Lagrange equation of \dLparal\ , we obtain
\begin{equation}
    \label{eq:delta_Bpar_equation_motion}
\delta \ddot{B}_{\parallel, \ell}(t) = - \left( \delta B_{\parallel, \ell}(t) + B_{0} \right ) \frac{4 V_{A}^{2}}{\ell_{\perp, 0}^{2}},
\end{equation}
where $V_{A} = B_{0}/\sqrt{4 \pi \rho}$ is the Alfv\'enic speed.

Initially, we compressed the perturbed volume perpendicular to \Bordered, then released it and allowed the compression to propagate. For the initial conditions, we considered that $u_{\perp, \ell}(t=0)=0$ and $\delta B_{\parallel, \ell}(t=0)=\delta B_{\parallel,\rm max}$. We might have initiated the fluid parcel at $\delta B_{\parallel, \ell}(t=0)=-\delta B_{\parallel,\rm max}$, but in that case, $u_{\perp, \ell}(t=0) \neq 0$. Solutions of Eq.~\ref{eq:delta_Bpar_equation_motion} are harmonic, but are valid only for early times because at later times, nonlinear interactions become important and energy is diffused. Below, we consider the scenario of energy diffusing due to the shock formation because we considered highly compressible fluids. Without loss of generality, we can consider any diffusive process.

From the jump conditions, we analytically obtained that when \MS\ $\gg 1$, an isothermal shock perpendicular to \Bordered\ forms when
\begin{equation}
    \label{eq:shock_condition}
    \delta B_{\parallel} \lesssim \frac{B_{0}}{2} \left( \mathcal{M}^{2}_{A} -1 \right)
.\end{equation}
Thus, in sub-Alfvénic turbulence, \MA~$< 1$, magnetized shocks form when \dBparal\ $< 0$, which  means that \dBparal\ will never perform a full harmonic cycle. Keeping the dominant term of the expansion of the harmonic solutions (Eq.~\ref{eq:delta_Bpar_equation_motion}), we derive that
\begin{equation}
    \label{eq:delta_B_parall_solution}
    \delta B_{\parallel, \ell} (t) \approx \delta B_{\parallel, \max} - 2 B_{0} \frac{V_{A}^{2}}{\ell_{\perp, 0}^{2}} t^{2}.
\end{equation}
The above solution through Eq.~\ref{eq:u_perp_generalized} yields 
\begin{equation}
    \label{eq:u_perp_solution}
    u_{\perp, \ell}(t) \approx \frac{2 V^{2}_{A}}{\ell_{\perp, 0}} t.
\end{equation}
From Eqs.~\ref{eq:delta_B_parall_solution} and \ref{eq:u_perp_solution}, we obtain that as the magnetic field of the perturbed volume decreases, \uperpl\ increases. When the shock is formed, the perturbed volume instantaneously bounces off its environment, which acts as a pressure wall \citep{Basu_2009}. At the post-shock phase, the motion is reversed and the coherent volume will start contracting until until \dBparall\ reaches a value of $+\delta B_{\parallel, \max, p}$. The post-shock solutions are obtained from Eq.~\ref{eq:delta_Bpar_equation_motion} with initial conditions $u_{p}(t=0) > 0$ and $\delta B_{\parallel, p} (t=0)<0$, where the subscript $p$ denotes post-shock quantities. At the post-shock phase, the solution of \dBparal\ is\begin{equation}
    \delta B_{\parallel, \ell} (t) \approx - \delta B_{\parallel, p} + u_{p}(t=0)t - 2 B_{0} \frac{V_{A}^{2}}{\ell_{\perp, 0}^{2}} t^{2}.
\end{equation}

At the post-shock phase, the magnetic field increases until $+\delta B_{\parallel, \max, p}$, which is smaller than the initial magnetic field increase ($+\delta B_{\parallel, \max}$) of the pre-shock phase because energy has been dissipated by the shock \citep{park_2019_shocks, cho_2022_shocks}. When the perturbed volume reaches $+\delta B_{\parallel, \max, p}$, the velocity is zero, and the motion is reversed. Then, the volume starts expanding until it again forms a shock. Overall, the perturbed volume would perform damped oscillations until all the energy is dissipated \citep{Basu_2009, yang_2021}.

Fluids in nature are commonly assumed to be constantly perturbed until turbulence reaches a steady state \citep{krumholz_2010, kritsuk_2017, colman_2022}. Various driving mechanisms could maintain turbulent energy in nature \citep{eswaran_1988, mckee_1989, mac_low_1998, piontek_2007, elmegreen_2009, krumholz_2016, girichidis_2016, Shravan_TurbDriv_2020, olivier_2017, klessen_2010, elmegreen_scalo_2004}. In our model, turbulent driving is equivalent to adding externally kinetic energy to the perturbed volume, such that the initial velocity at the post-shock phase, $u_{p}(t=0)$, is sufficient to compress the perturbed volume until it reaches the maximum compression it had in the pre-shock phase, $\delta B_{\parallel, \max, p} \approx +\delta B_{\max, \parallel}$.

We considered an external driver, which ensured that \dBparal\ fluctuations, and hence energy, were maintained in a quasi-static state. In addition, we considered that the fluid is ergodic \citep{monin_ergodicity, galanti_2004_ergodicity}. For ergodic fluids, \dBparall\ are characterized by ballistic profiles, $\delta B_{\parallel, \ell} \propto t^{2}$, and as we argue below, they bounce between $+\delta B_{\parallel,\max}$ and $-\delta B_{\parallel,\max}$ within a characteristic timescale $\mathcal{T}_{\perp} \approx 4 \ell_{\perp, 0} V_{A}^{-1} \sqrt{\delta B_{\parallel,\max}/2 B_{0}}$.

When we initially compressed the magnetic field of the perturbed volume along \Bordered, then  \Lperp\ decreased due to Eq.~\ref{eq:db_parall_lperp}. This forced the surface of the environment of the perturbed volume to increase by equal amounts. Thus, the $+\delta B_{\parallel,\max}$ initial  increase of the magnetic field of the perturbed volume causes the magnetic field of the environment to decrease by $-\delta B_{\parallel,\max}$ due to flux freezing. If the fluid is ergodic, then different fluid parcels correspond to different oscillation phases of the target fluid parcel \citep{monin_ergodicity, galanti_2004_ergodicity}. Therefore, the $-\delta B_{\parallel,\max}$ of the environment corresponds to the maximum decrease in magnetic field strength of the target volume. Nonlinear effects can break the symmetry between $+\delta B_{\parallel, \max}$ and $-\delta B_{\parallel, \max}$, but ergodicity is only weakly broken when \Bordered\ $\neq 0$ \citep{shebalin_2013_ergodicity}.

The perturbed volume would spend most of its time in the compressed state because the velocity is minimum there. On the other hand, the velocity of the fluid parcel is maximum when $\delta B_{\parallel, \ell} < 0$, and hence the fluid parcel would spend minimum time there. As a result, the majority of fluid parcels at a given time would be compressed ($\delta B_{\parallel, \ell} > 0$) due to ergodicity, which is verified by numerical simulations \citep{beattie_2022_energy_balance}.

\section{Solutions of \dLperp}
\label{sec:solutions_dlperp}

From the Euler-Lagrange equation of \dLperp\ , we obtain 
\begin{equation} 
    \label{eq:delta_Bperp_equation_motion}
    \delta \ddot{B}_{\perp, \ell}(t) \delta B_{\perp, \ell}(t) - 2\delta \dot{B}_{\perp, \ell}^{2} (t)+ \frac{\delta B_{\perp, \ell}^{6}(t)}{4 \pi \rho C^{2}} = 0.
\end{equation}
For $|\vec{B}_{0}| \gg |\delta \vec{B}|$, the sixth-order term above can be neglected, and then the solutions are straightforward. The total pressure of the fluid exerted by \dBperp\ is transferred to parallel motions (Eq.~\ref{eq:u_parallel_generalized}), hence $\rho u_{\parallel, \max}^{2}/2 = \delta B_{\perp, \max}^{2}/(8\pi)$. We derive the following solutions:
\begin{equation}
    \label{eq:deltaB_perp_solution}
    \delta B_{ \perp, \ell} (t) \approx \frac{f B_{0}}{1 \pm t/\mathcal{T}_{\parallel}},~ u_{\parallel, \ell}(t) \approx \pm f V_{A},
\end{equation}
where $\ell_{\parallel}(t=0) = \ell_{\parallel, 0}$, $f=\delta B_{\perp, \max}/B_{0} \ll 1$, and $\mathcal{T}_{\parallel}=fV_{A}^{-1} \ell_{\parallel, 0}$. In the above equations, the signs depend on the initial conditions. Initially, we considered that $\delta B_{\perp, \ell}(t=0) = \delta B_{\perp, \max}$, and $u_{\parallel, \ell}(t=0) = u_{\perp, \max}$, which leads to positive signs. 

If the initial velocity along \Bordered\ were zero, then both \uparallell\ and \dBperpl\ would remain static. The coupling of parallel and perpendicular motions (Eq.~\ref{eq:u_parallel_generalized}) would induce parallel motions when $\delta \dot{B}_{\parallel, \ell} \neq 0$, even if $u_{\parallel, \ell}(0)=0$. However, because we have neglected the coupling of motions, we initiated \uparallell\ from the initial conditions. 

From Eq.~\ref{eq:db_perp_lparal}, we obtain that the free streaming of the perturbed volume causes \Lparallel\ to expand or contract as
\begin{equation}
    \label{eq:L_parallel_solution}
    \ell_{\parallel} (t) \approx \ell_{\parallel, 0} \left(1 \pm \frac{t}{\mathcal{T}_{\parallel}} \right).
\end{equation}
As the target fluid parcel expands, its environment along the \Bordered\ axis contracts, provided that the fluid has fixed boundaries. Due to the expansion of the target volume, the initial velocity of the environment would be $- u_{\parallel, \max}$, which causes a negative sign in the denominator of Eq.~\ref{eq:deltaB_perp_solution}, and hence \dBperpl\ increases in the environment. On the other hand, \dBperpl\ in the target volume stops increasing when $t = \mathcal{T}_{\parallel}$ because \dBperpl\ in the environment becomes infinite. In sub-Alfvénic flows, $|\vec{B}_{0}| \gg |\delta \vec{B}_{\perp}|$, so that this infinity should be treated as an asymptotic behaviour of \dBperpl: there is a physical limit above which \dBperpl\ cannot grow. After $\mathcal{T}_{\parallel}$, the motion is reversed and the environment starts expanding along \Bordered, causing the target volume to contract with \dBperpl\ growing as \footnote{This solution is obtained by considering that the initial conditions in the reversed motion of the fluid parcel are $\delta B_{\perp, \ell}(0)=\delta B_{\perp, \max}/2$, $u_{\parallel, \ell}(0)=-u_{\parallel, \max}$, and $\ell_{\parallel}(0) = 2 \ell_{\parallel, 0}$. These values correspond to the solutions of Eqs.~\ref{eq:deltaB_perp_solution} and \ref{eq:L_parallel_solution} for $t=T_{\parallel}$.} $\delta B_{\perp, \ell}(t) \approx fB_{0}/(2-fV_{A}\ell_{\parallel, 0}^{-1}t)$ until it reaches $\delta B_{\perp, \max}$.

\end{appendix}

\end{document}